\documentclass[useAMS,usenatbib,usedcolumn]{mn2e}
\usepackage{epsfig,graphicx}

\begin{document}
\title{The host galaxies of type Ia supernovae at $z=0.6$}
\author[D.~Farrah et al]{D.~Farrah$^1$, W.P.S~Meikle$^1$,
D.~Clements$^1$, M.~Rowan-Robinson$^1$, S.~Mattila$^1$\\
$^1$Astrophysics Group, Blackett Laboratory, Imperial College, Prince
Consort Road, London SW7 2BW, UK\\} \date{2002 May 30} \pubyear{2002}
\volume{000} \pagerange{1} \twocolumn

\maketitle \label{firstpage}
 
\begin{abstract} 
We examine the host galaxies of high redshift type~Ia supernovae
(SNe~Ia) using archival I and R band data from the Hubble Space
Telescope.  The SNe~Ia host galaxies show a wide variety of
morphologies, including undisturbed ellipticals, spirals and disturbed
systems. SNe~Ia are also found over a wide range of projected
distances from the host galaxy centres, ranging from $3$kpc to
$\sim30$kpc.  For a sample of 22 SNe~Ia at $\langle z\rangle = 0.6$, 
$\sim70\%$ are found in spiral galaxies and $\sim30\%$ are found in elliptical
systems, similar to the proportions observed locally.  Including data
from \citet{ell}, we find no significant difference in the average
light-curve-shape-corrected $M_{B}^{peak}$ for high-z SNe~Ia between
spirals and ellipticals.  These results are consistent with
predictions based on the locally-derived understanding of SNe~Ia
physics and the influence of progenitor mass and metallicity.  We also
construct colour maps for two host galaxies and find that both show a
non-uniform colour structure with typical variations of rest-frame
$B-V\sim0.5$.  This
is most plausibly attributed to the presence of, and variation in dust
extinction in these galaxies.  Moreover, we find no evidence that the
SNe~Ia are preferentially found in outer regions ($>10$~kpc) of the host
galaxies where extinction would be low.  This suggests that the range
of host galaxy extinctions of SNe~Ia at $z\sim0.6$ should be
comparable to those of local SNe~Ia.  Although observational bias cannot 
be completely ruled out, this appears to be in conflict with the finding 
of low extinction for SNe~Ia found in the high-$z$ supernova search studies.
\end{abstract}

\begin{keywords}
 supernovae: general --  galaxies: high-redshift -- galaxies: stellar content 
-- cosmological parameters
\end{keywords}

\section{Introduction}
Observations of type~Ia supernovae (SNe~Ia) at high redshift has led
to the astonishing deduction that the universal expansion is
accelerating \citep{rie1,per1}. A key factor in this discovery has
been the painstaking calibration of the relation between the SNe~Ia
luminosity and the behaviour of the light curves
\citep{phi0,ham,rpk,per0,tri,phi,sah}.  However, the size of the
effect (about 0.25~mags) which indicates acceleration is several times
smaller than the intrinsic range of luminosity in local SNe~Ia {\it
i.e.} the correction that has to be made using the calibration
relation is large.  Given that the calibration relation is derived
empirically, and that the physics of SNe~Ia is only partially
understood, it is therefore important to exploit every opportunity to
test for intrinsic differences in the SNe or in their environments
between the local and high-z universe.  In this letter we consider the
rates, luminosities and extinction of SNe~Ia at $z\sim0.6$.

The current consensus on the diversity of SNe~Ia is that it arises
primarily from the mass of $^{56}$Ni produced in the explosion, with a
higher relative $^{56}$Ni producing a more luminous and more slowly
declining lightcurve.  Close-binary evolutionary calculations
\citep{ume1} of possible progenitors indicate that the C/O ratio is
lower for higher mass CO cores.  It has been suggested, although not 
demonstrated, that a smaller C/O ratio in the progenitor white dwarf 
yields less $^{56}$Ni and so a lower luminosity \citep{hwt,ume1}. Thus, 
CO white dwarf progenitors arising from higher mass main-sequence stars 
may yield less luminous SNe~Ia.  Metallicity can have a significant 
effect via the white dwarf wind which affects the minimum CO white 
dwarf mass that can accrete as far as the Chandrasekhar Limit \citep{kob}. 
A high metallicity implies a higher white dwarf wind.  This allows 
white dwarfs with a lower CO mass to reach the Chandrasekhar Limit and 
hence produce brighter SNe~Ia. Thus, there will be a tendency for a 
larger number of, and more luminous SNe~Ia to occur in high metallicity 
populations and vice versa.  It is predicted that metallicity in spirals 
will decrease with increasing redshift \citep{cal}.  Such a decrease 
is not expected to occur in ellipticals since they have had essentially 
no star formation over this look-back time.  Consequently we might 
expect that, relative to ellipticals, SNe~Ia in high-$z$ spirals should 
exhibit reduced luminosities and lower rates. 

Conversely, modelling of the total SNe~Ia rate (per $10^{10}$
$L_{B\odot}$) with redshift predicts little change in the total 
SNe~Ia rate with increasing redshift up to $z\sim1.6$ \citep{ktn}. 
This prediction has been shown to be true up to at least $z\sim0.55$
by independent observations \citep{pai}. Moreover, \citet{ktn} also 
model the variation in SNe~Ia rate with redshift for host galaxies of
different morphological types, and find that significant changes are
anticipated only beyond z$\sim$1.6.  At lower redshifts, the absolute
and relative rates are predicted to remain roughly the same as the
local values for spirals and ellipticals.
 
Another important issue in current studies of high-$z$ SNe~Ia is that
of extinction.  Both supernova search teams correct for galactic
extinction but find that very few of their supernovae show a residual
colour excess, implying negligible extinction in the host galaxies.
No significant extinction is found in 14 out of 16 SNe~Ia discovered
by \citet{rie1} and in 40 out of 42 SNe~Ia discovered by \citet{per1}.
\citet{rie1} also argue against the presence of neutral extinction.
They point out that the observed dispersion in MLCS-derived distances
is substantially smaller than would be expected if grey dust
extinction were the main cause of the lower brightness of high-$z$
SNe~Ia.  However, \citet{lei} has highlighted inconsistencies in the
treatment of host galaxy extinctions between the two SNe~Ia search
teams.  Moreover, \citet{mrr} has shown that by employing a consistent
treatment of extinction and excluding those SNe~Ia observed only after
maximum light, a positive $\Lambda$ is only required at the
$\sim3\sigma$ level.

The apparent lack of host galaxy extinction is, perhaps, surprising
for two reasons. Firstly, local SNe~Ia are observed with a range of
host galaxy extinctions spanning $0 < E(B-V) < 2.0$, with most lying
in the range $0 < E(B-V) < 0.5$ \citep{phi,mkm}.  Secondly, the
increasing gas content in galaxies with increasing redshift means
that, contrary to the findings of the SN search teams, we might expect
mean extinction levels to {\it increase} with redshift due to
increased dust opacities caused by the higher gas column densities
\citep{cal}.  However, it is possible that the apparent lack of
extinction at high-z is due to observational bias.  SNe~Ia behind a
dust screen will be fainter, and thus would be selected against in
searches near the magnitude limit.  In addition, follow-up spectroscopy
favours those SNe~Ia that are a large distance from the host galaxy
centres in order to minimise contamination from host galaxy light.
These SNe~Ia may be expected to have less extinction than those found
near the host galaxy centres.

\begin{table}
\caption{High redshift type Ia supernovae host galaxies
\label{sne1ahosts}}
\begin{tabular}{@{}lccccc}
\hline
\hline
Name            & RA         & Dec                      & $z$  & Host   & $r$$^{a}$  \\
                & hh mm ss   & \degr\ \arcmin\ \arcsec\ &      &        & Kpc  \\        
\hline
\hline
1997ce$^{b}$ & 17 07 48.3 & +44 01 26.2              & 0.44 & Spiral    & 3.1  \\  
1997cj$^{b}$ & 12 37 04.3 & +62 26 24.9              & 0.50 & E/S0      & 4.8  \\
1997ek$^{c}$ & 04 56 11.6 & -03 41 26.0              & 0.86 & Spiral    & 6.3  \\
1997eq$^{c}$ & 04 58 56.3 & -03 59 29.4              & 0.54 & Sab       & 13.8 \\
1997es$^{c}$ & 08 18 40.7 & +03 13 36.5              & 0.65 & Disturbed & 3.0  \\
1998ba$^{d}$ & 13 43 36.9 & +02 19 30.6              & 0.43 & Spiral    & 3.8  \\
1998bi$^{d}$ & 13 47 44.7 & +02 20 57.2              & 0.75 & Spiral    & 6.2  \\
1998J$^{e}$  & 09 31 10.5 & -04 45 36.5              & 0.83 & Spiral?   & 3.3  \\
1998M$^{e}$  & 11 33 44.4 & +04 05 13.4              & 0.63 & E/S0      & 28.9 \\
1999U$^{f}$  & 09 26 43.0 & -05 37 57.8              & 0.50 & E/S0      & 8.6  \\
2000dz$^{g}$ & 23 30 41.4 & +00 18 42.7              & 0.50 & Spiral    & 3.5  \\
2000ea$^{g}$ & 02 09 54.0 & -05 28 17.8              & 0.42 & Spiral    & 9.0  \\
2000ec$^{g}$ & 02 11 32.0 & -04 13 56.1              & 0.47 & E/S0      & 20.7 \\
2000ee$^{g}$ & 02 27 34.5 & +01 11 49.4              & 0.47 & Spiral    & 4.8  \\
2000eg$^{g}$ & 02 30 21.1 & +01 03 48.5              & 0.54 & Spiral    & 6.4  \\
\hline
\hline
\end{tabular}

\medskip

J2000 coordinates and redshifts are taken from the relevant IAU circulars. 
$^{a}$Projected distance in kpc of the SNe~Ia from the host galaxy 
centres. $^{b}$\citet{rie1}, $^{c}$\citet{nug}, $^{d}$\citet{gre}, 
$^{e}$\citet{gar} $^{f}$\citet{gar2} $^{g}$\citet{sch}

\end{table}

In this letter we use HST archive data, and previously published HST
observations, to test whether or not SNe~Ia rates and luminosities at
$z\sim0.6$ show a different distribution with host morphology.  We
also examine the issue of extinction via the radial distribution of
the SNe, and by making use of colour maps of two of the high-z host
galaxies.  Sample selection, observations and data analysis are
described in \S2. Results are presented in \S3, and discussion and
conclusions are given in \S4. We adopt $H_{0}=65$ km s$^{-1}$
Mpc$^{-1}$, $\Omega_{0}=0.3$ and $\Lambda=0.7$.

\section{The Sample}

Our sample, presented in Table \ref{sne1ahosts}, comprises 15 SNe~Ia
events discovered in the course of the two current high-$z$ SNe~Ia
search programmes, and also observed by the HST.  All except one of
the supernovae were firmly classified as type Ia events on the basis
of their spectra, with SN 1997es being classified as a probable type
Ia. The mean redshift of the sample is $\langle$$z\rangle = 0.57$ with
a range spanning $0.42 < z < 0.86$.

Galaxy images were obtained from the Wide Field Planetary Camera 2
(WFPC2) archive, in the F675W and F814W filters which correspond to
the Cousins R and I bands respectively.  Each supernova was observed
at several epochs spaced a few weeks apart in order to sample the
light curves, with the deeper images being in the F814W band.  For
each event, the F814W band data from these epochs were co-added to
form a single deep image of the SN and host galaxy.  Most of the
supernovae were centred on the WF3 chip, with a few centred on the
Planetary Camera. Data taken at each epoch were combined into a single
frame using the STSDAS task CRREJ, which also performs sky
subtraction.  The images for each epoch were then aligned using the
IRAF task CROSSDRIZ and combined into a single exposure.  The final
coadded images have effective exposure times of between 1000s and
2400s. This method of construction means that, although the SNe~Ia are
visible, their magnitudes cannot be measured here since the combined
images comprise exposures taken over a range of epochs.  It does
however give very deep images of the host galaxies, allowing effective
classification of the host galaxy types.

To investigate the dust content of the host galaxies, we constructed
colour maps using the F675W and F814W images.  The images in each
filter were first scaled to a common exposure time, and then
deconvolved with PSFs using the Richardson-Lucy algorithm \citep{luc}
as implemented in the IRAF task L{\sevensize UCY}. The TinyTIM v6.0
software \citep{kri} was used to generate PSFs for the appropriate
filter and chip positions.  Any colour structure not due to colour
differences in the images will be due to two effects. The first is
differences in the effective PSF's of the images achieved by the
initial deconvolutions. To quantify this difference we performed
simulations by constructing colour maps of stars visible in both
images. It was found that the maps of these stars had a featureless
colour structure across the diffraction disk. Colour structure due to
PSF mismatch is therefore negligible compared to noise in the original
images or large angle scattering within the WFPC2 optics. The second
effect is that of deconvolution artifacts. It was found that
deconvolution artifacts only became visible several iterations after
convergence between the original image and the deconvolved image had
been achieved.  As we halted the deconvolution immediately upon
attaining convergence, the effect of artifacts on the achieved colour
structure is insignificant.  We found that, due to the pixel scale of
WFPC2, resolved colour structure was only attainable for galaxies
which extended over more than $1\arcsec$ and which had deep imaging in
both filters.  Thus, the colourmap study was restricted to just two
galaxies, the hosts of SN~2000ea and SN~2000eg.

\begin{figure*}
\begin{minipage}{170mm}
\epsfig{figure=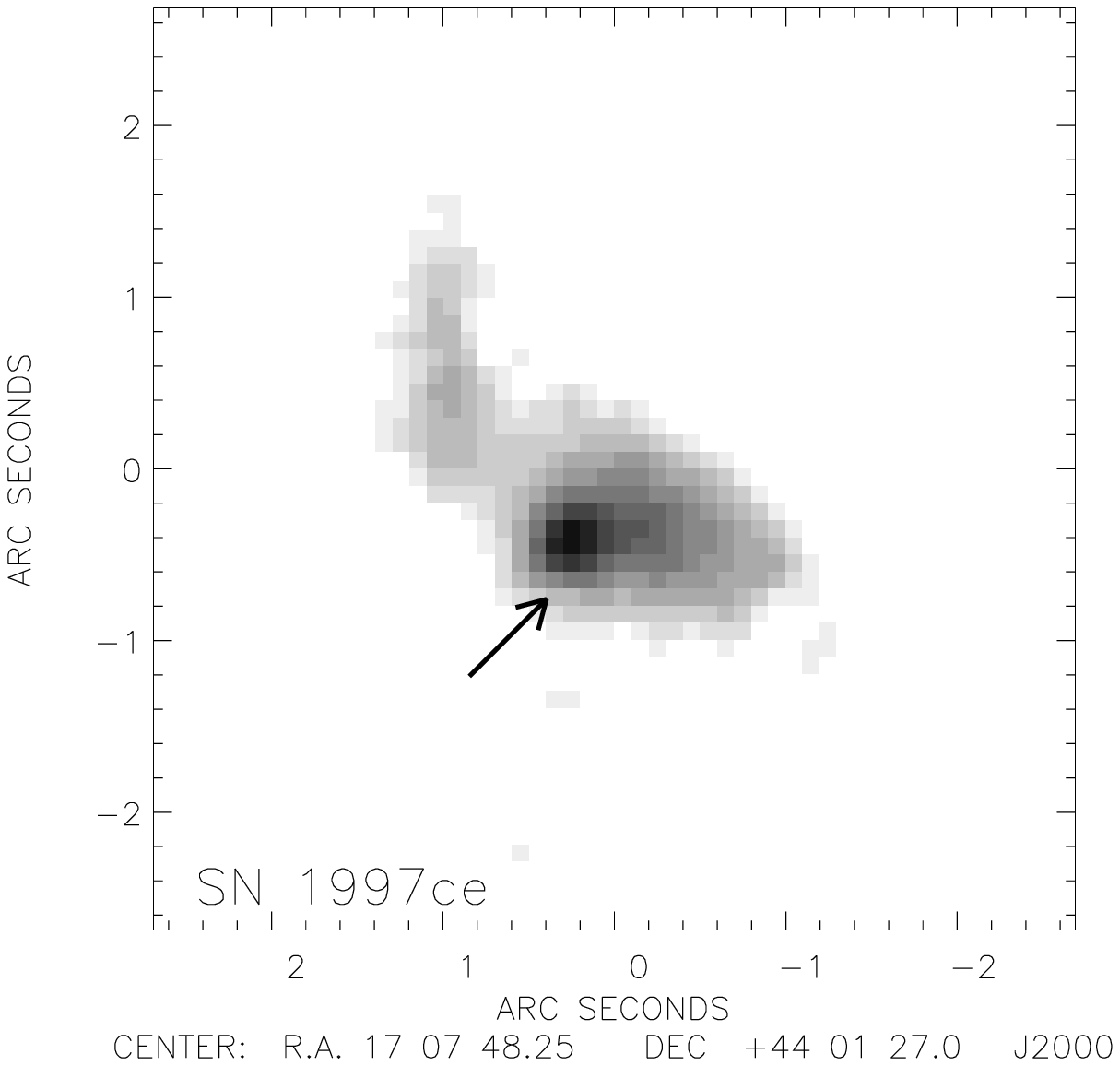,width=50mm}
\epsfig{figure=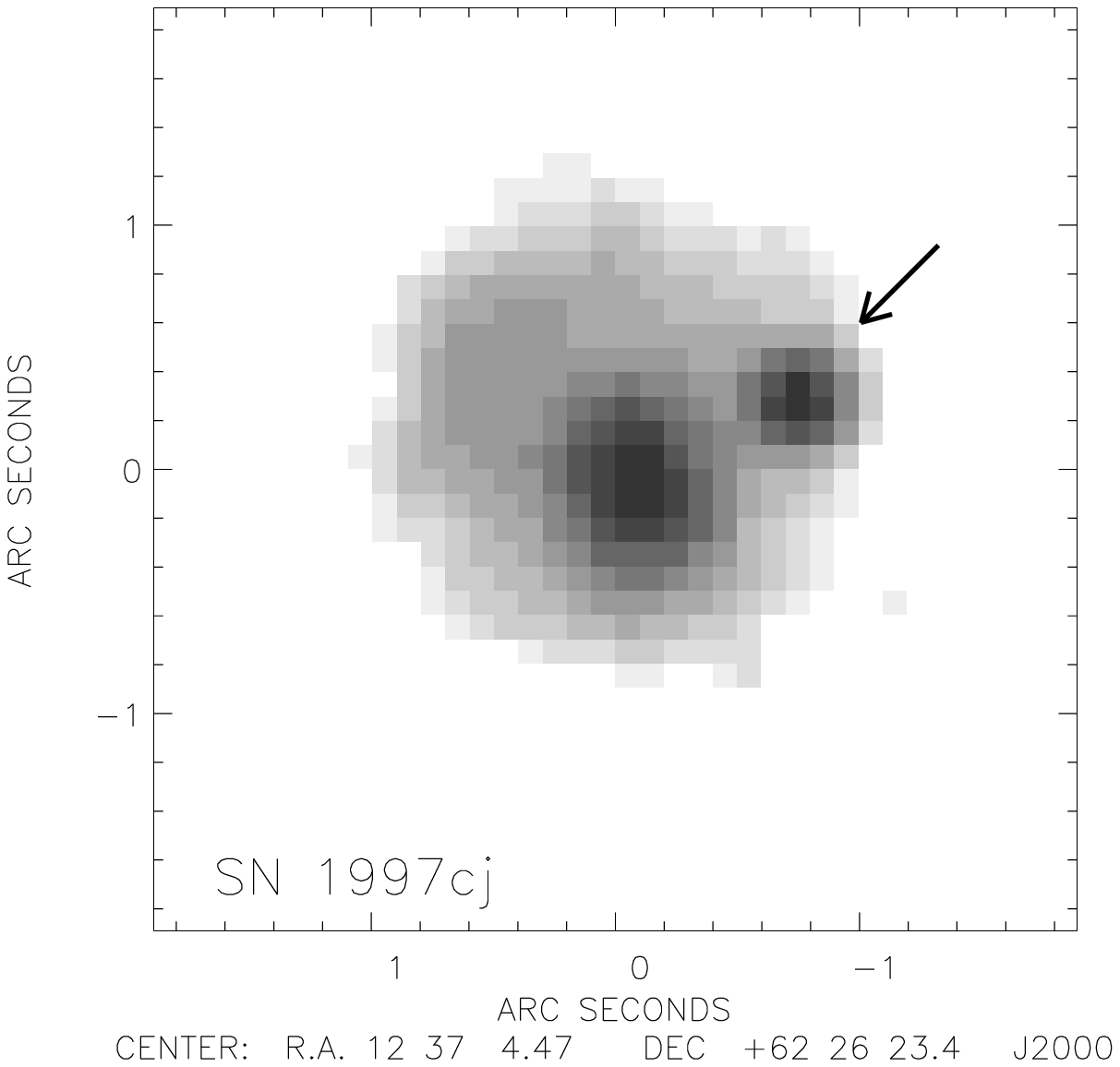,width=50mm}
\epsfig{figure=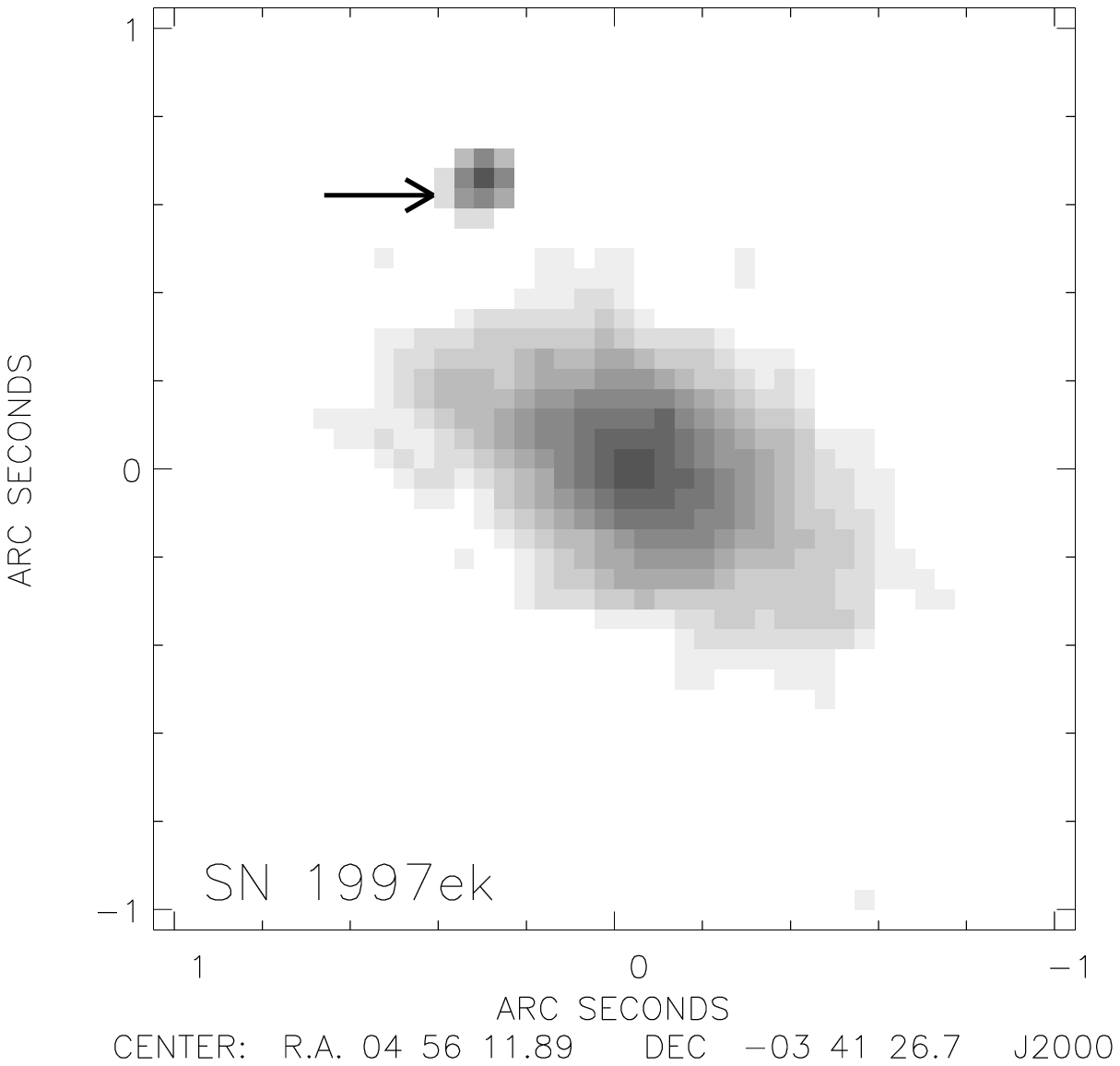,width=50mm}
\end{minipage}
\begin{minipage}{170mm}
\epsfig{figure=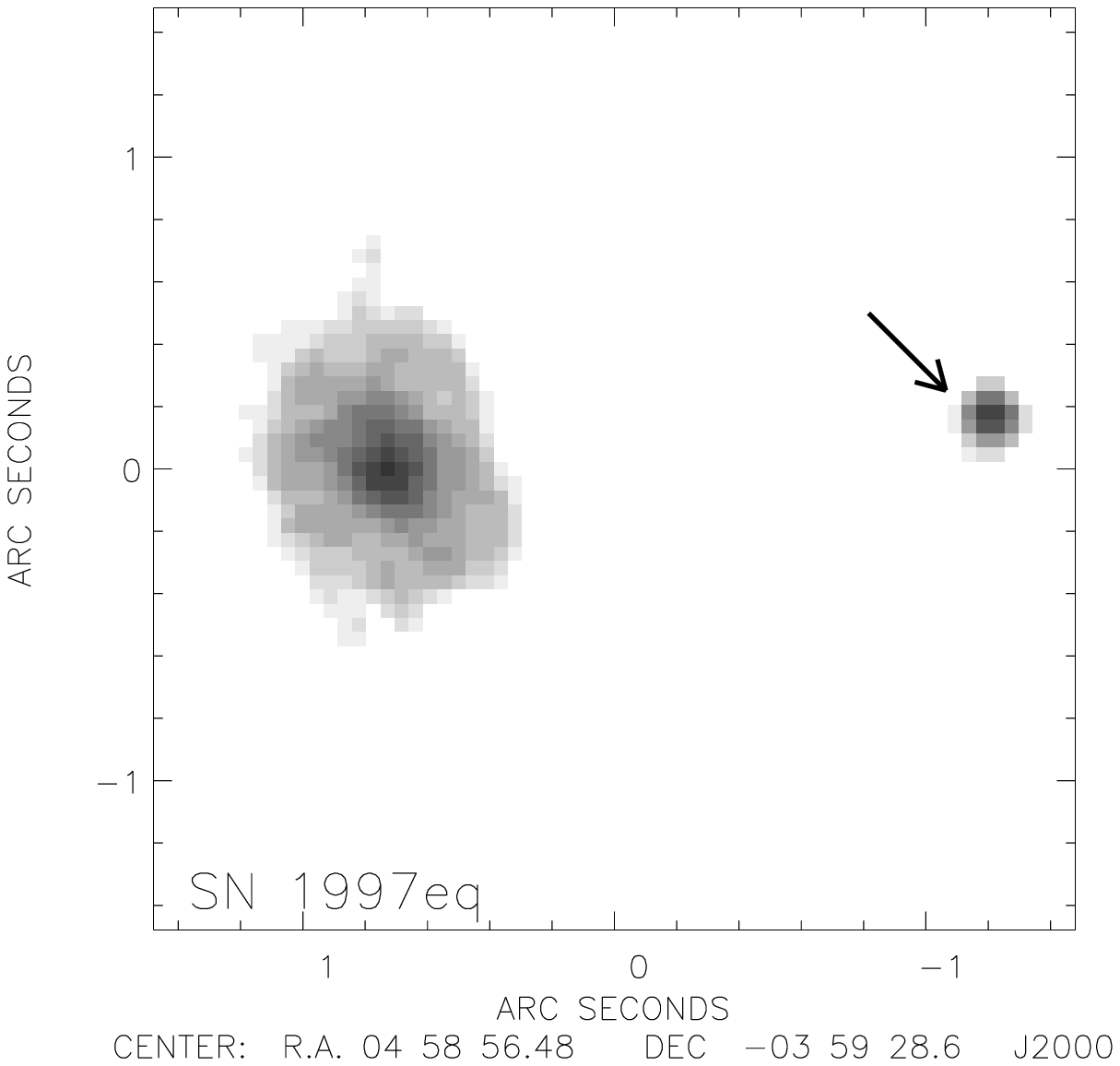,width=50mm}
\epsfig{figure=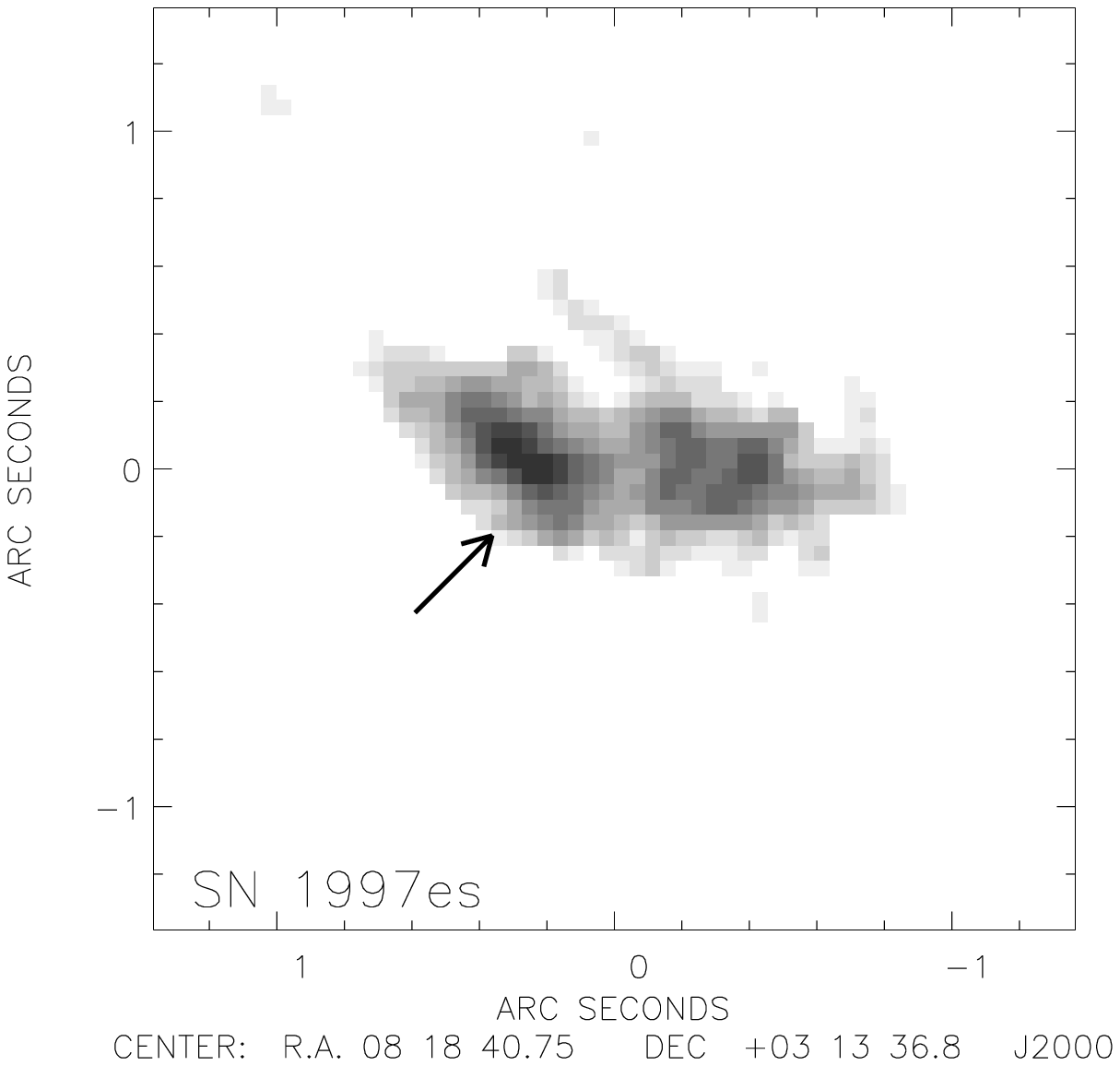,width=50mm}
\epsfig{figure=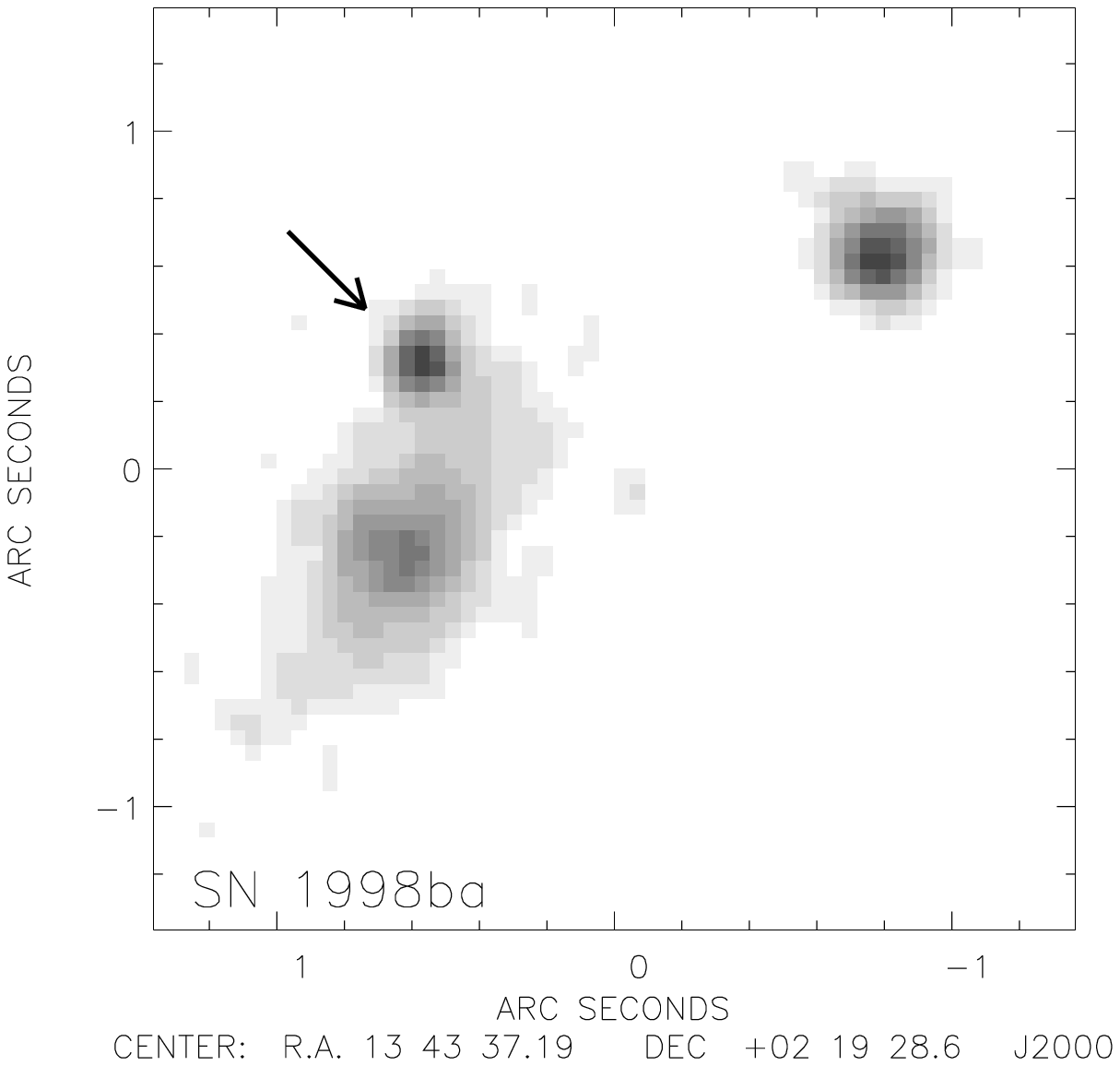,width=50mm}
\end{minipage}
\begin{minipage}{170mm}
\epsfig{figure=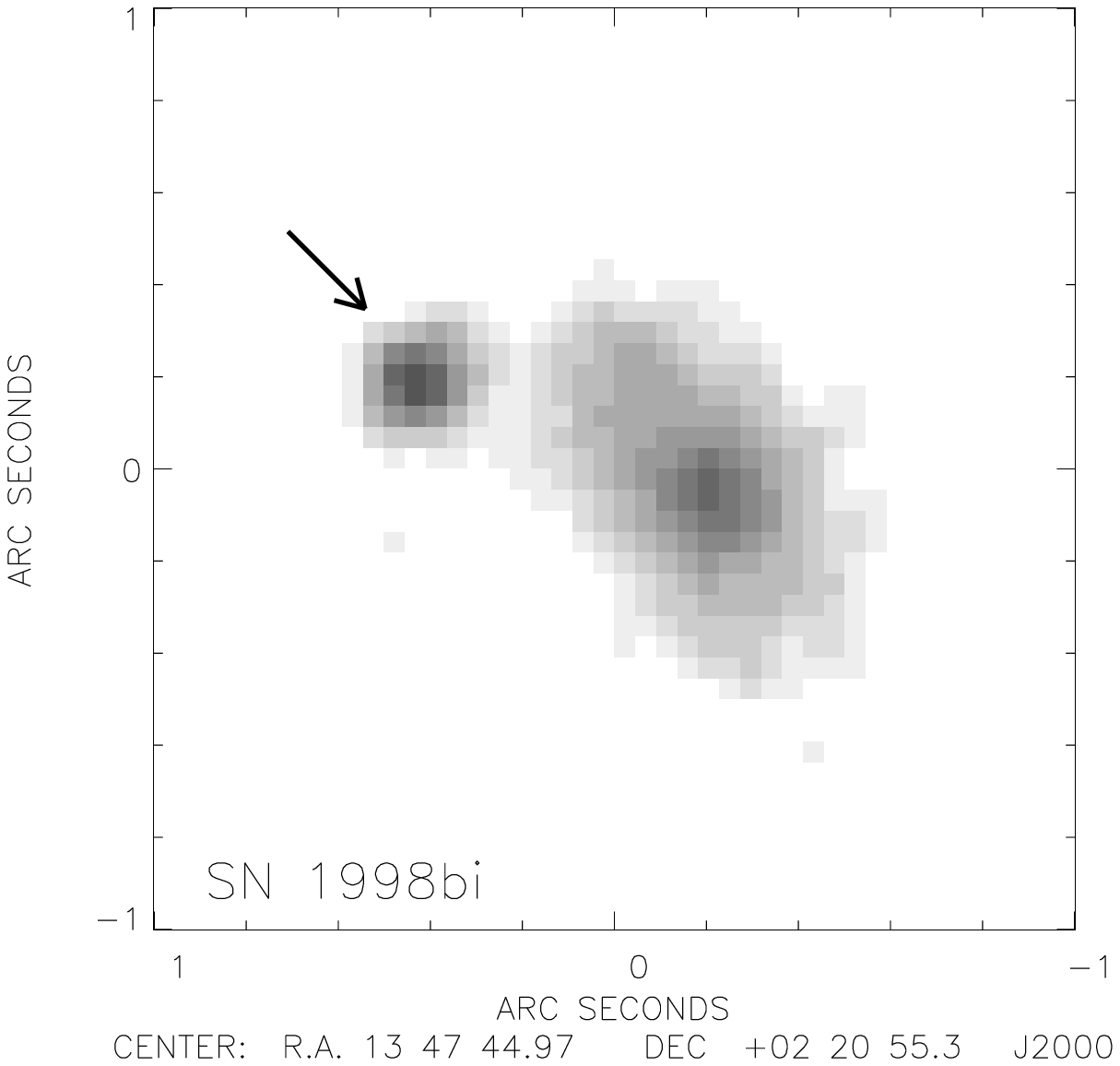,width=50mm}
\epsfig{figure=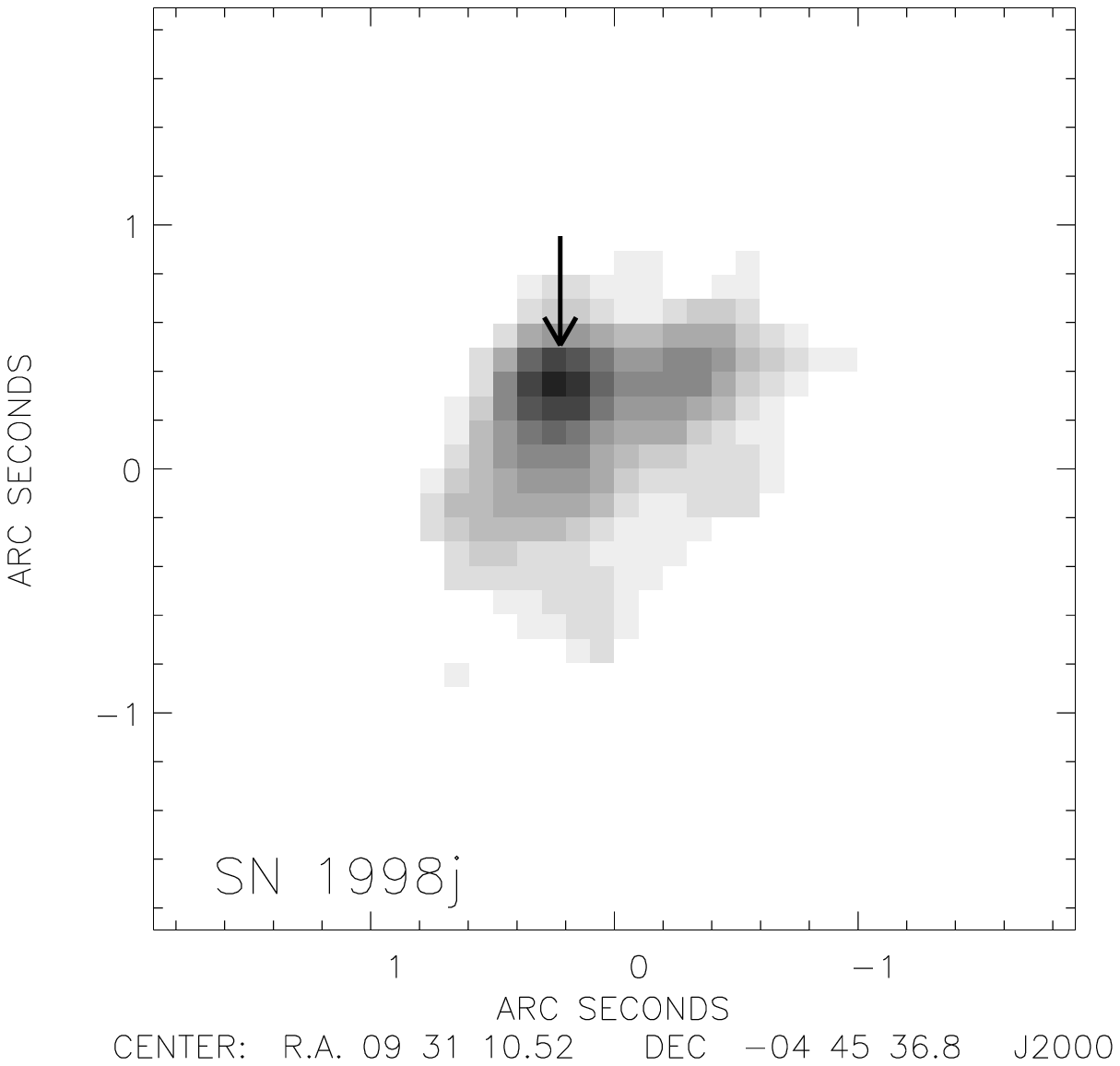,width=50mm}
\epsfig{figure=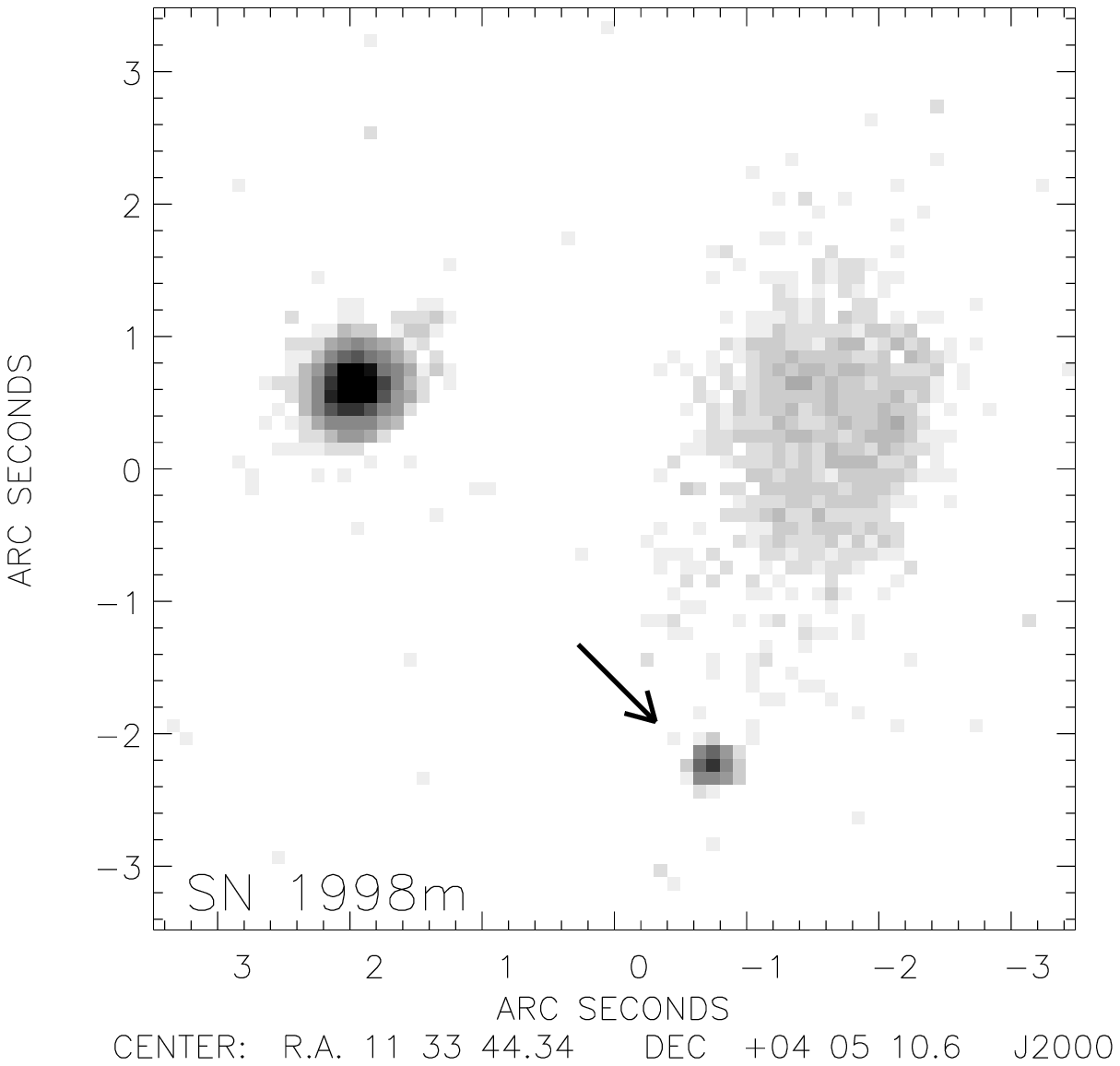,width=50mm}
\end{minipage}
\begin{minipage}{170mm}
\epsfig{figure=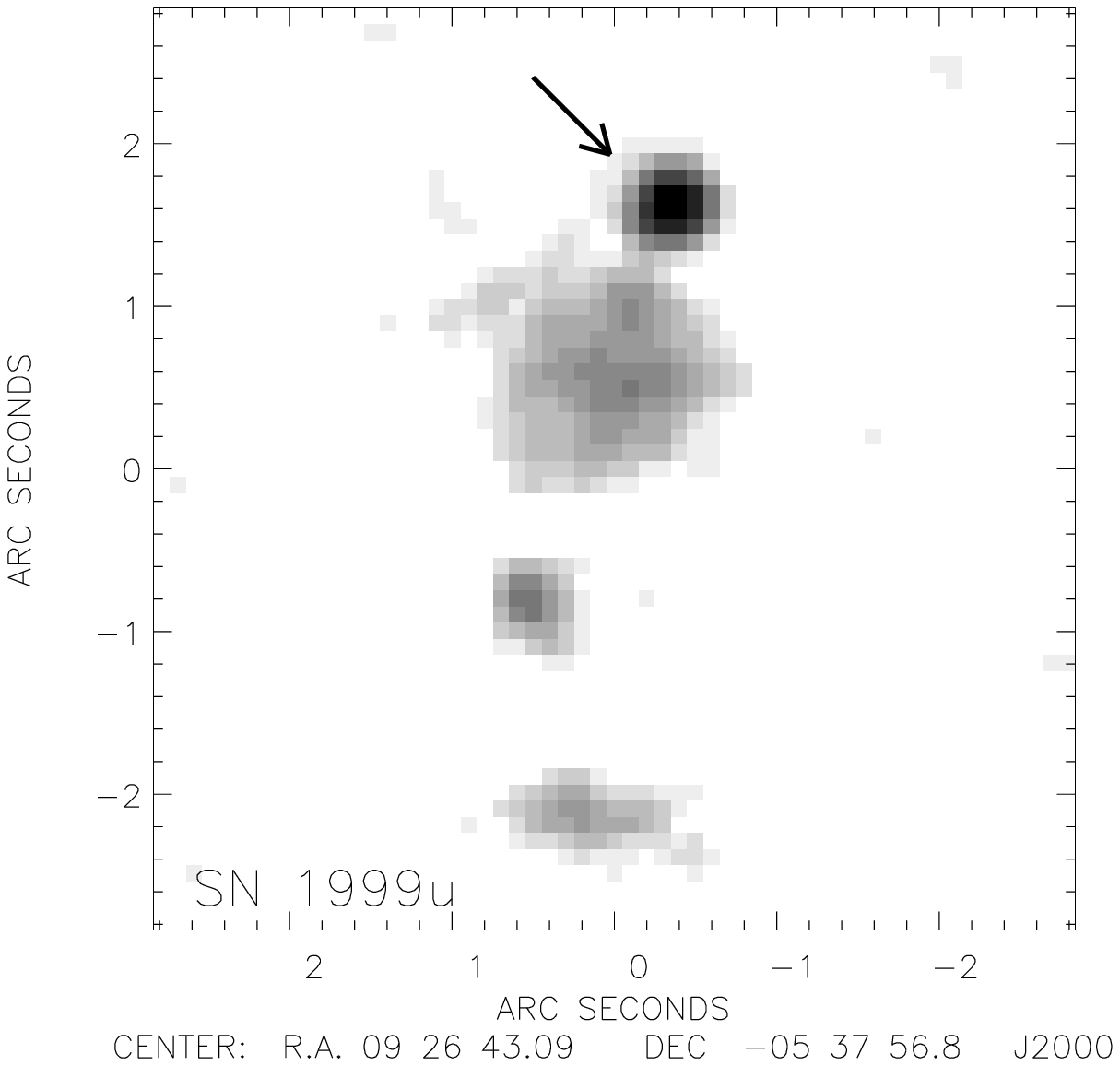,width=50mm}
\epsfig{figure=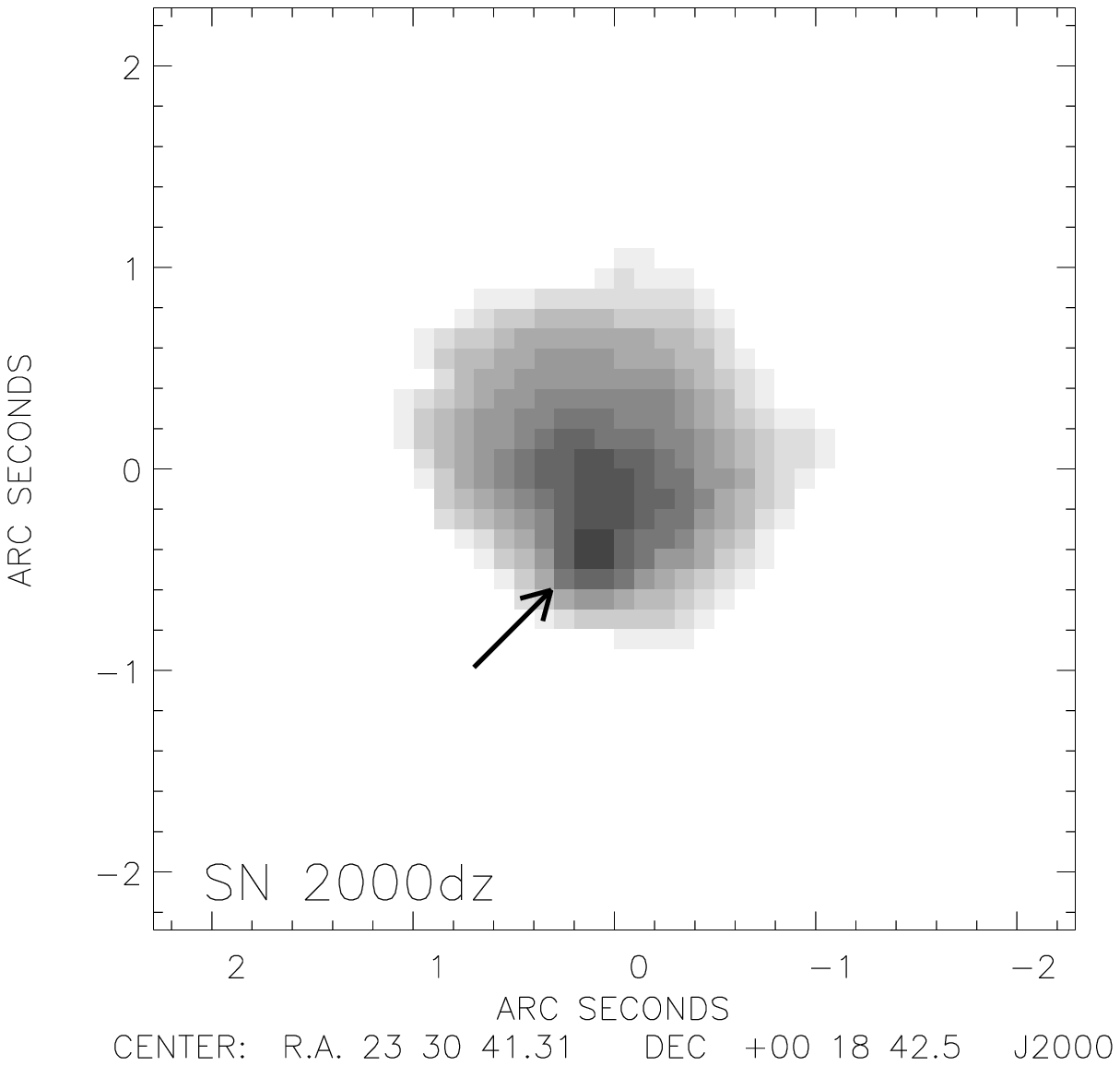,width=50mm}
\epsfig{figure=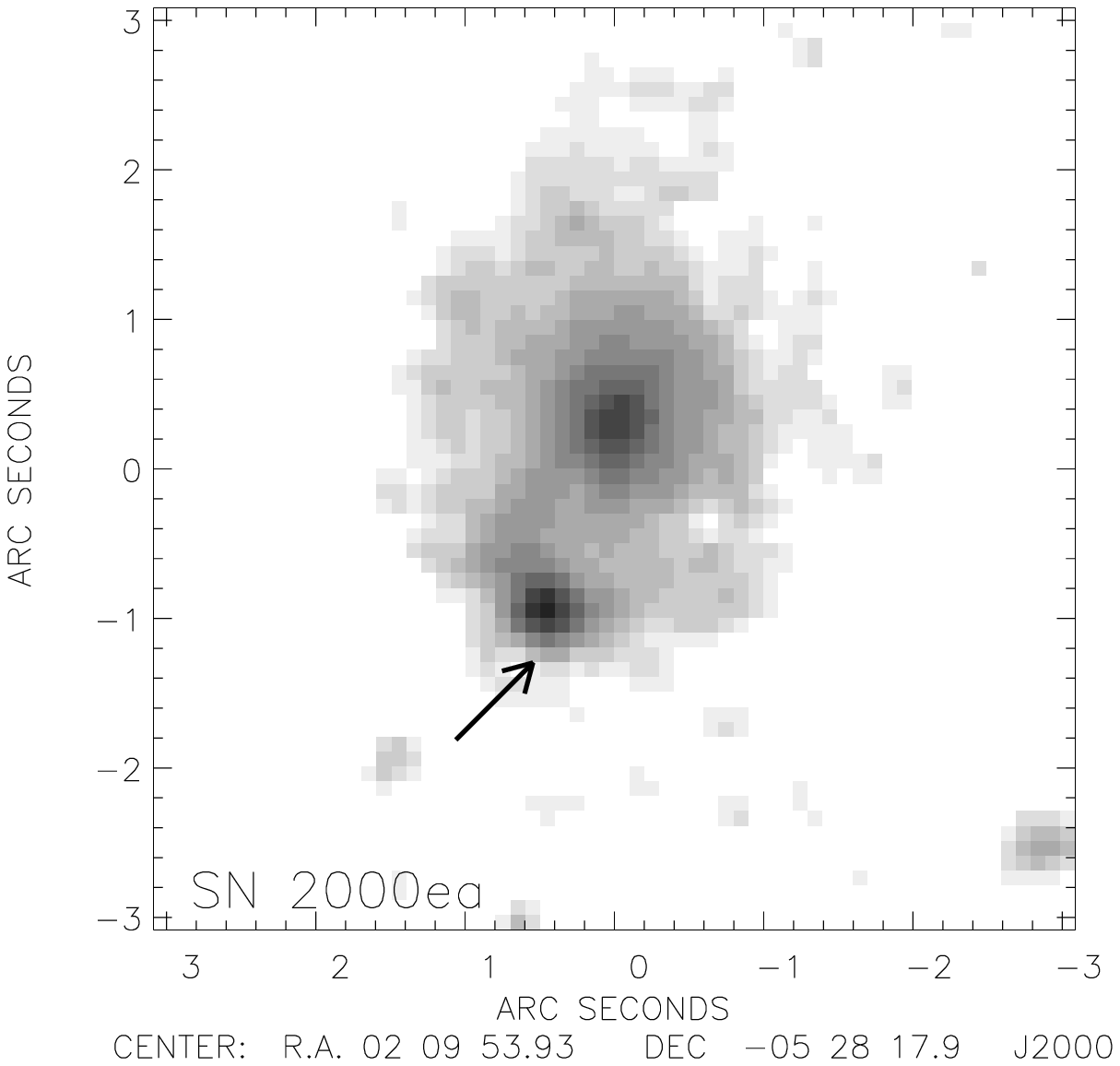,width=50mm}
\end{minipage}
\begin{minipage}{170mm}
\epsfig{figure=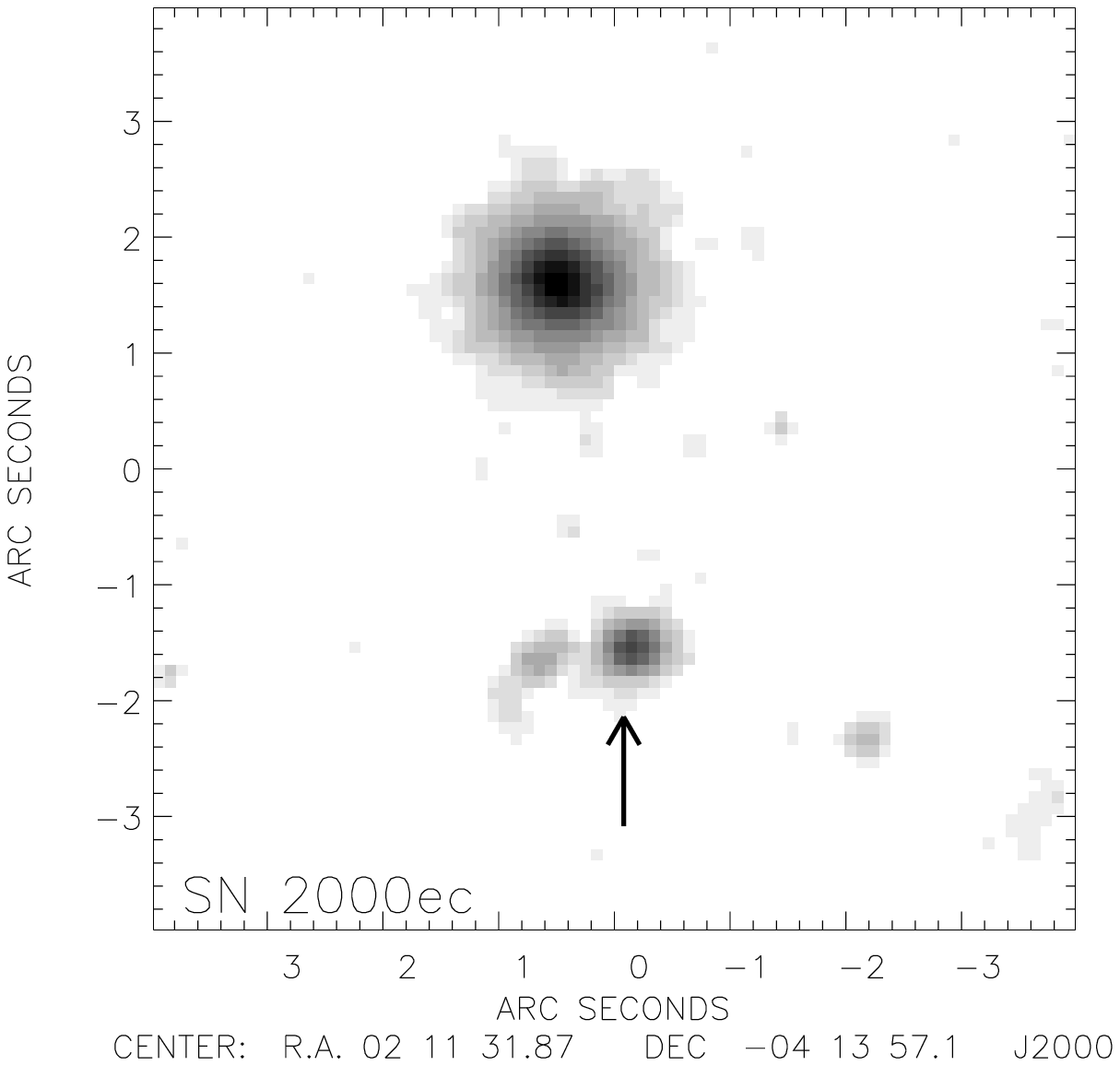,width=50mm}
\epsfig{figure=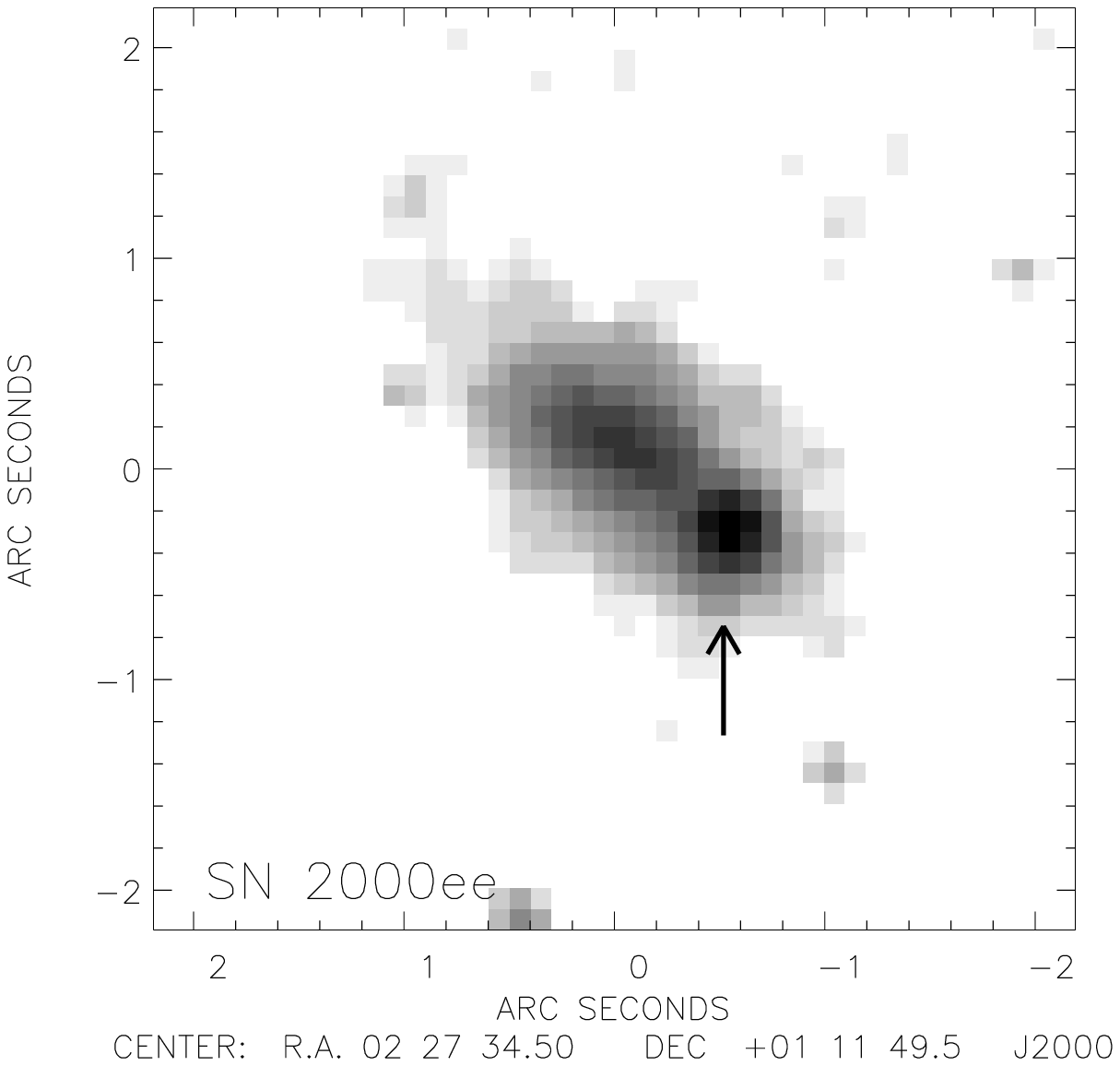,width=50mm}
\epsfig{figure=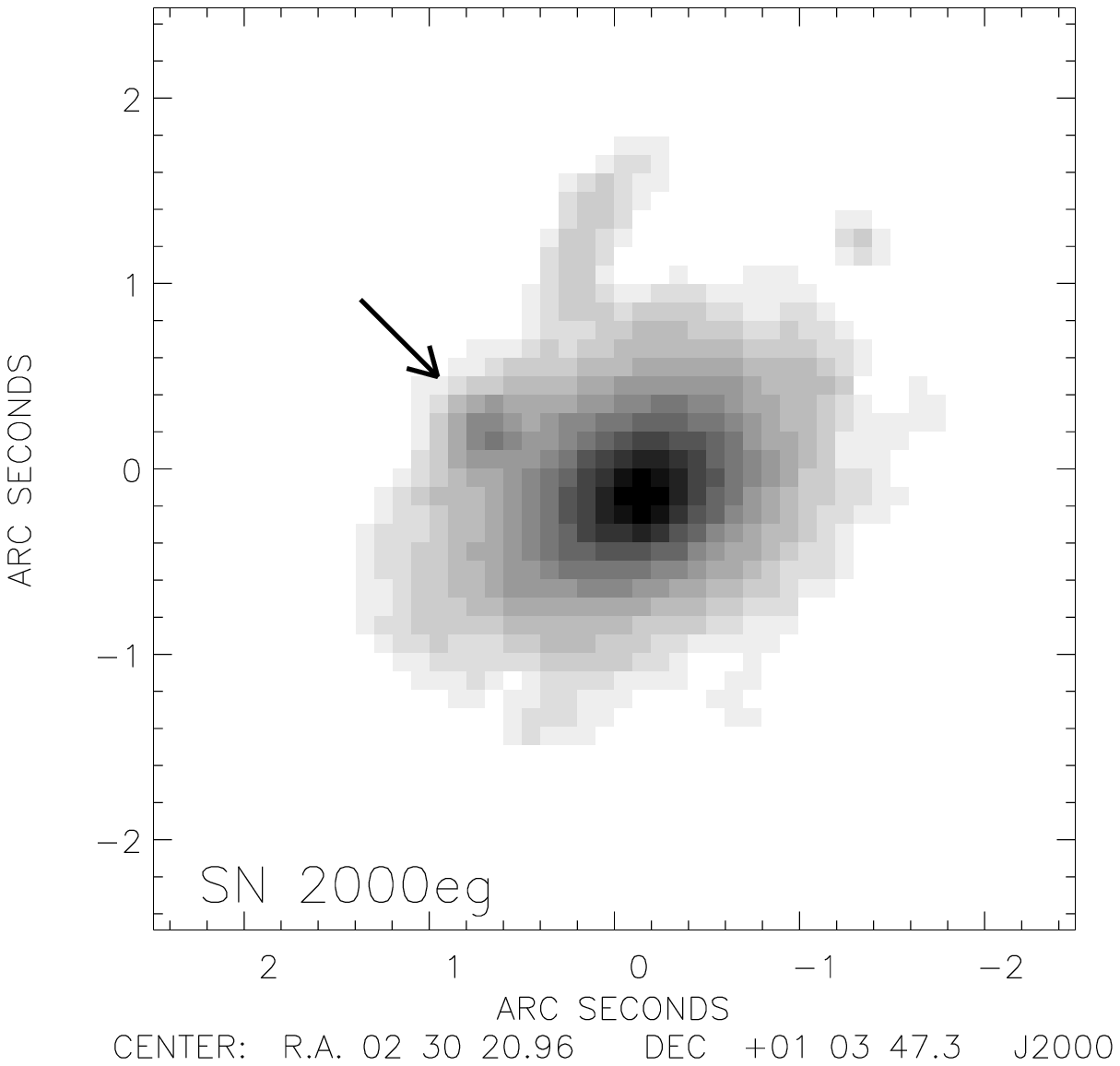,width=50mm}
\end{minipage}

\caption{ F814W host galaxy images for the 15 supernovae in our
sample. The supernovae are indicated by arrows.
\label{sn_images} }
\end{figure*}

\begin{table}
\caption{$z\sim0.5$ SNe Ia hosts from \citet{ell}
\label{ellishosts}}
\begin{tabular}{@{}lccccc}
\hline
\hline
Name            & RA         & Dec                      & $z$  & Host   & $M_{B}^{peak}$ \\
                & hh mm ss   & \degr\ \arcmin\ \arcsec\ &      &        &                \\        
\hline
\hline
1994an          & 22 44 18.9 & +00 06 48.7              & 0.38 & Scd    & -19.11         \\ 
1995ar          & 01 01 20.4 & +04 18 33.8              & 0.47 & Sa     & -18.95         \\ 
1995as          & 01 01 35.3 & +04 26 14.8              & 0.50 & Sbc    & -18.73         \\ 
1995ax          & 02 26 25.8 & +00 48 44.2              & 0.62 & E/S0   & -19.81         \\ 
1995az          & 04 40 33.6 & -05 30 03.6              & 0.45 & Scd    & -19.66         \\ 
1997eq          & 04 58 56.3 & -03 59 29.4              & 0.54 & Sab    &  --            \\  
1997F           & 04 55 14.3 & -05 51 44.8              & 0.58 & SBbc   & -19.37         \\ 
1997H           & 04 59 36.6 & -03 09 34.6              & 0.53 & E/S0   & -19.44         \\ 
\hline
\hline
\end{tabular}

\medskip

J2000 coordinates are quoted. Peak B band magnitudes, given in the
rest-frame of the supernova, are taken from \citet{per1} and include
corrections using the stretch-factor method. SN1997eq is present in
both the \citet{ell} sample and our sample in Table \ref{sne1ahosts}.

\end{table}

\section{Results}

The final F814W host galaxy images are presented in Figure
\ref{sn_images} and morphological classifications are given in Table
\ref{sne1ahosts}.  Spectra for several of the supernovae in
our sample have been published \citep{rie1,lso}, although no spectra of the host
galaxies themselves have been published. Where possible we have used 
the contamination from host galaxy light in the published SNe~Ia spectra to
refine the morphological classifications further. We note however that 
classification of host galaxy type based on the observed morphology only
is still robust in distinguishing between elliptical and spiral
systems.  However, the images on their own are insufficiently detailed to be used
to divide the spiral systems into subtypes (e.g. Sbc, SBbc etc.).

The sample exhibits morphologies ranging from pure ellipticals
(e.g. the hosts of SN~1997cj, SN~2000ec) to systems with
clear signs of spiral structure (the hosts of SN~1997ce, SN~1997ek,
SN~1998bi). The host galaxy of SN~1997es appears to be disturbed.
Overall, four hosts have a morphology consistent with an E/S0
classification, while the remaining eleven either exhibit spiral
structure or are disturbed.  A wide range of SN distances from the
galaxy centres is also found, ranging from $3$kpc (SN~1997es,
SN~2000ee) to $\sim30$kpc (SN~1998M. Note that the object directly
above the supernova is a foreground galaxy.) The projected distances
are presented in Table \ref{sne1ahosts}.

\begin{figure*}
\begin{minipage}{170mm}
\epsfig{figure=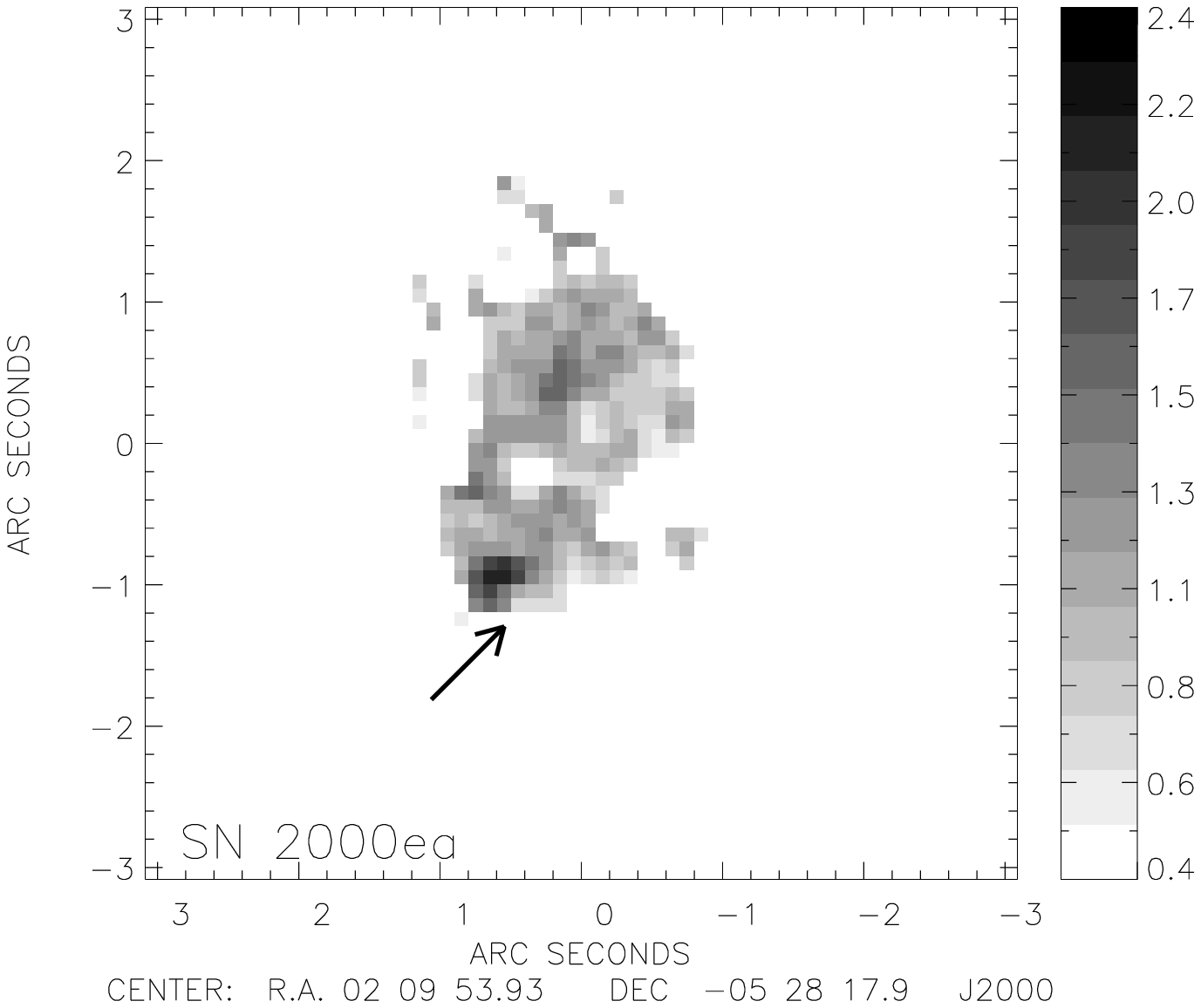,width=70mm}
\epsfig{figure=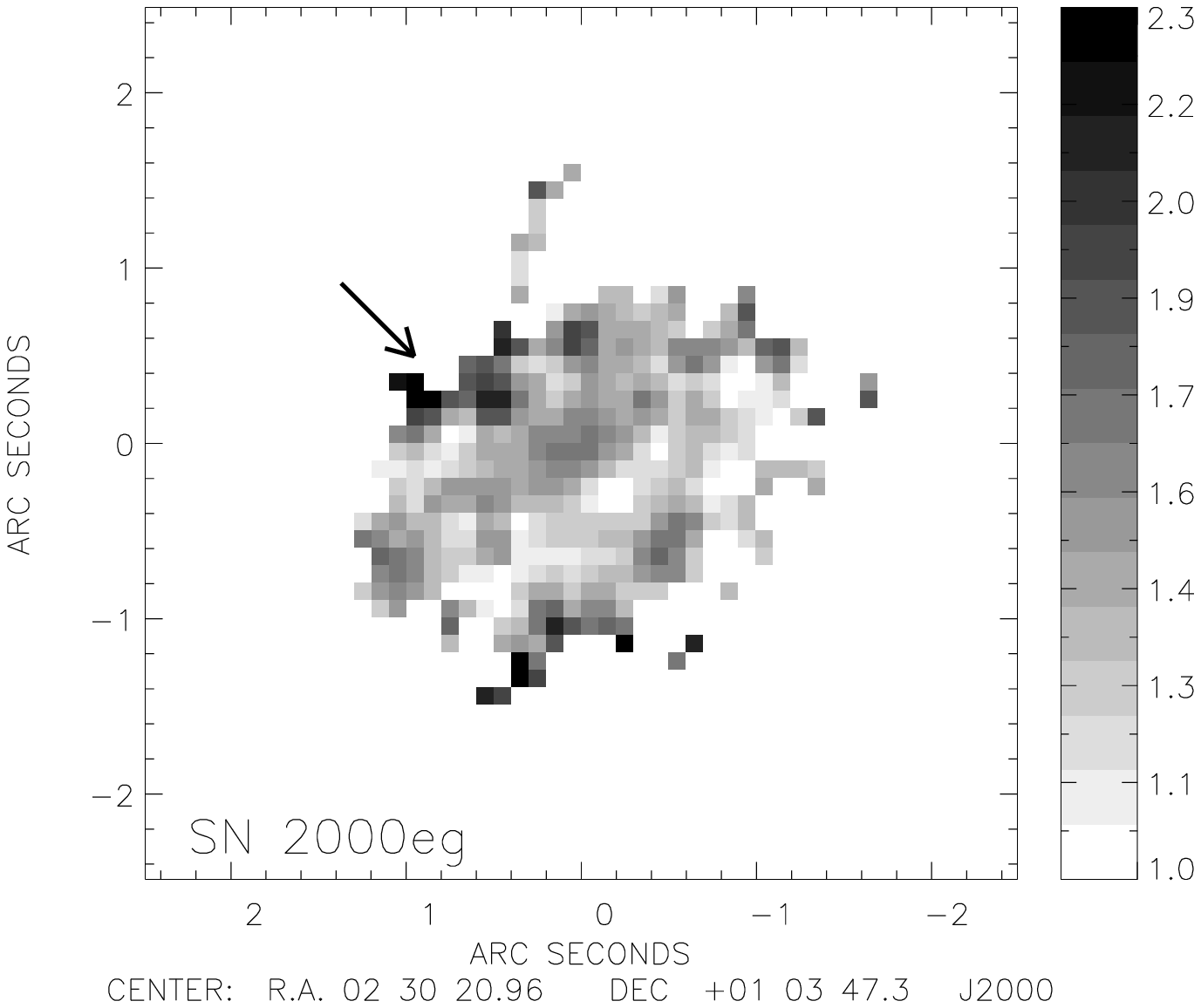,width=70mm}
\end{minipage}
\caption{ Host galaxy colourmaps for SN~2000ea and
SN~2000eg. An arrow indicates the position of the supernova. The vertical 
bar on the right hand side of each image gives the $m_{675}-m_{814}$ colour, 
which in the rest-frame of the objects corresponds closely to $B-V$ colour.  
\label{sn_colmaps} }
\end{figure*}

The colour maps for the host galaxies of SN~2000ea and SN~2000eg are
presented in Figure \ref{sn_colmaps}. Sky background regions in both maps 
have been
masked out.  In the rest frame of the galaxies, the colour corresponds
closely to $B-V$.  We note that, owing to the way in which the maps
were constructed, they cannot be used to accurately measure the supernova
colours.  Large-scale, irregular colour variation of order $B-V = 0.5$
is seen across both galaxies, with extreme differences as high as $B-V
= 1.4$.

\section{Discussion and Conclusions}
\subsection{Supernova rates}
To improve the statistics we added seven more SNe~Ia host galaxies at
a similar redshift, imaged by \citet{ell} using STIS on HST. These
galaxies are presented in Table \ref{ellishosts}. Two of these are
classified as E/S0 and five as spirals. An eighth galaxy from
\citet{ell}, the host of SN~1997eq, is common to both samples. We
agree with \citet{ell} that it is a spiral type.  Thus, for a combined
sample of 22 SNe~Ia host galaxies, $30\%$ are E/S0 systems and $70\%$
are spirals or disturbed systems. These proportions are similar to that observed locally by
\citet{cet}. These authors also find a ratio of about $30\%$ to $70\%$
when their results are expressed as SN rate per unit bolometric
luminosity of the host galaxy.  Adopting the same conversion between
SNe~Ia counts per galaxy type and SNe~Ia rates per unit galaxy
luminosity as \citet{cet}, we infer that the relative SNe~Ia rate in
spirals and ellipticals has not changed between $z\sim0.6$ and the
local universe.  This is consistent with the prediction of \citet{ktn}
that the relative rates should be unchanged in going from z=0.6 to the
local universe.

\subsection{Supernova peak magnitudes}
Peak rest-frame B band magnitudes for the SNe in our sample have so
far been published for only two objects \citep{rie1}; SN 1997ce
($M_{B}^{peak}$=--19.26) and SN 1997cj (--19.25). These are the
magnitudes obtained following application of MLCS correction.
However, stretch-factor-corrected B band peak magnitudes for 7 out of
the 8 host galaxies presented in \citet{ell} have been published in
\citet{per1}.  These magnitudes are listed in Table
\ref{ellishosts}. The total of 9 SNe~Ia comprises 3 in ellipticals and
6 in spirals. Although the two teams observe their supernovae using 
different filter sets, the light curve fitting methods employed by both
teams give stretch factor corrected magnitudes in the same filter, namely B band in the 
rest-frame of the supernovae. We thus intercompare published 
magnitudes from both teams. The mean peak B~band magnitude for the 
SNe~Ia in the 6 spiral systems is $-19.22^{-0.13}_{+0.14}$. For the 3 SNe~Ia in
elliptical systems the mean peak B~band magnitude is
$-19.53^{-0.15}_{+0.19}$.  Thus, we find no significant difference in
the average light-curve-shape-corrected $M_{B}^{peak}$ for high-z
SNe~Ia between spirals and ellipticals. We thus find that light-curve shape 
correction methods \citep{sch0} appear to be valid for SNe~Ia at 
$z\sim0.6$ in all host galaxy types. 

\subsection{Extinction} 
The irregular colour structure in two of the host galaxies (Figure
\ref{sn_colmaps}) is most plausibly interpreted as being due to
variation in a significant dust extinction (this also supports their
classification as spirals). We also note that simulations of host
galaxy extinctions and radial distributions of supernovae \citep{hbd}
have shown that SNe~Ia within $10$kpc of the centres of spiral
galaxies will be observationally dimmer and have a larger
magnitude dispersion than SNe~Ia in the outer regions, although part
of the observed dispersion will probably be due to projection
effects. Inspection of Figure \ref{sn_images} shows that the SNe~Ia
are found over a range of projected distances from the host galaxy
centre. These distances are listed in Table \ref{sne1ahosts}. It is
possible that the real distances from the host galaxy centres are in
fact all $>10$kpc and that the observed distances are simply due to
projection effects, however we argue against this for two
reasons. Firstly, 10 out of the 15 SNe in our sample have projected
distances from the host galaxy nuclei of $<7$kpc, which argues against
a distribution in a 'shell' of $>10$kpc in radius. Secondly, the
distances from galaxy centres for local SNe~Ia range from $\sim1$kpc
to $\sim30$kpc, and there is no compelling physical reason why the 
distribution of galactocentric distances of SNe~Ia at z=0.6 should differ 
from the distribution observed locally.

Overall, our results are consistent with a similar range of real
separations of SNe~Ia from the centres of their host galaxies as is
observed locally.  We find no evidence to suggest that SNe~Ia at
$z\sim0.6$ are found preferentially far out in the host galaxies where
extinction levels might be expected to be small. Moreover, current 
observational and theoretical evidence does not favour any significant 
changes in typical galaxy size at $z\sim0.6$ relative to that observed 
locally \citep{lil,boi}. Consequently we can also state that there is 
no difference in the distribution with galaxy scalelength between local 
SNe~Ia and those at $z\sim0.6$. We are led to suspect that a significant 
fraction of the 22 events may be subject to some extinction.  Given that 
dustier environments at higher redshifts are also to be expected \citep{eal,fox}, 
the general lack of extinction found in high-$z$ SNe~Ia discovered by 
\citet{rie1} and \citet{per1} remains puzzling. We do however note the following 
two caveats. Firstly, the possibility of some observational bias ({\it cf.} Section 1) 
operating within a region closer than $\sim$10~kpc is not ruled out. Secondly, 
details about the SNe~Ia in our two ``dusty'' spirals (Figure \ref{sn_colmaps}) 
have yet to be published, and so we are unable to comment directly on the degree of
extinction that these events might have encountered.

In summary, we find that, at $z\sim0.6$, the observed relative rates
and relative luminosities of type Ia supernovae in elliptical and
spiral galaxies are consistent with predictions based on the
locally-derived understanding of SNe~Ia physics and the influence of
progenitor mass and metallicity \citep{ume1,ktn}.  We find no reason to
question this understanding.  Unless some level of   
observational bias is present, there is however still some difficulty with the
apparently low observed extinction towards SNe~Ia at high-z. 

\section{Acknowledgments}
We thank Bruno Leibundgut for helpful discussions.  We would like to
thank the referee for very helpful comments.  The data were obtained
from the Multimission Archive at the Space Telescope Science Institute
(MAST). STScI is operated by the Association of Universities for
Research in Astronomy, Inc., under NASA contract NAS5-26555. Support
for MAST for non-HST data is provided by the NASA Office of Space
Science via grant NAG5-7584 and by other grants and contracts. The
NASA/IPAC Extragalactic Database (NED) is operated by the Jet
Propulsion Laboratory under contract with NASA. The Digitized Sky
Surveys were produced at the Space Telescope Science Institute under
U.S. Government grant NAG W-2166. The images of these surveys are
based on photographic data obtained using the Oschin Schmidt Telescope
on Palomar Mountain and the UK Schmidt Telescope. This work was in
part supported by PPARC (grant number GR/K98728). Financial support
was provided to S.M. by Osk. Huttusen S\"a\"ati\"o

\label{lastpage}


\begin{thebibliography}{}

\bibitem[Baade(1938)]{baa}
Baade W., 1938, ApJ, 88, 285

\bibitem[Boissier \& Prantzos(2001)]{boi}
Boissier S., Prantzos N., 2001, MNRAS, 325, 321

\bibitem[Branch, Romanishin \& Baron(1996)]{brb}
Branch D., Romanishin W., Baron E., 1996, ApJ, 467, 473

\bibitem[Calzetti \& Heckman(1999)]{cal}
Calzetti D., Heckman T. M., 1999, ApJ, 519, 27

\bibitem[Cappellaro, Evans \& Turatto(1999)]{cet}
Cappellaro E., Evans R., Turatto M., 1999, A\&A, 351, 459

\bibitem[Coil et al(2000)]{coi}
Coil A., et al, 2000, ApJ, 544L 111

\bibitem[Dahlen \& Fransson(1999)]{dah}
Dahlen T., Fransson C., 1999, A\&A, 350, 349

\bibitem[de Bernardis et al(2000)]{deb}
de Bernardis P., et al, 2000, Nature, 404, 955

\bibitem[Eales \& Edmunds(1996)]{eal}
Eales S. A., Edmunds M. G., 1996, MNRAS, 280, 1167

\bibitem[Ellis \& Sullivan(2001)]{ell}
Ellis R., Sullivan M., 2001, IAU Symposium 201, New Cosmological Data 
and the Values of the Fundamental Parameters (eds. A.
Lasenby and A. Wilkinson)

\bibitem[Evans \& Wilkinson(2000)]{evw}
Evans N. W., Wilkinson M. I., 2000, MNRAS, 316, 929

\bibitem[Fox et al(2002)]{fox}
Fox M. J., et al, 2002, MNRAS, 331, 839

\bibitem[Garnavich et al(1998)]{gar}
Garnavich P., et al, 1998, IAUC, 6819, 1

\bibitem[Garnavich et al(1999)]{gar2}
Garnavich P., et al, 1999, IAUC, 7097, 1

\bibitem[Green et al(1998)]{gre}
Green D., et al, 1998, IAUC, 6881, 1

\bibitem[Hamuy et al(1995)]{ham0}
Hamuy M., Phillips M. M., Maza J., Suntzeff N. B., Schommer R. A., Aviles R.,
1995, AJ, 109, 1

\bibitem[Hamuy et al(1996)]{ham}
Hamuy M., Phillips M. M., Suntzeff N. B.,
Schommer R. A., Maza J., Aviles R., 1996, AJ, 112, 2391

\bibitem[Hatano, Branch \& Deaton(1998)]{hbd}
Hatano K., Branch D., Deaton J., 1998, ApJ, 502, 177

\bibitem[Hoeflich, Wheeler \& Thielemann(1998)]{hwt}
Hoeflich P., Wheeler J. C., Thielemann F. K., 1998, ApJ, 495, 617

\bibitem[Howell(2001)]{how}
Howell D. A., 2001, ApJ, 554L, 193

\bibitem[Kobayashi et al(1998)]{kob}
Kobayashi C., Tsujimoto T., Nomoto K., Hachisu I.,
Kato M., 1998, ApJ, 503L, 155

\bibitem[Kobayashi, Tsujimoto \& Nomoto(2000)]{ktn}
Kobayashi C., Tsujimoto T., Nomoto K., 2000, ApJ, 539, 26

\bibitem[Krist(1994)]{kri}
Krist J., 1995, ADASS, 77, 349

\bibitem[Leibundgut(2001)]{lei}
Leibundgut B., 2001, ARA\&A, 39, 67

\bibitem[Leibundgut \& Sollerman(2001)]{lso}
Leibundgut B., Sollerman J., 2001, Europhysics News, 32, 4

\bibitem[Lilly et al(1998)]{lil}
Lilly S., et al, 1998, ApJ, 500, 75

\bibitem[Lucy(1974)]{luc}
Lucy L. B., 1974, AJ, 79, 745

\bibitem[Meikle \& Mattila(2002)]{mkm}
Meikle W. P. S., Mattila S., 2002, IAUC, 7911

\bibitem[Navasardyan et al(2001)]{nav}
Navasardyan H., Petrosian A. R., Turatto M., Cappellaro E., Boulesteix J., 
2001, MNRAS, 328, 1181

\bibitem[Nugent et al(1998)]{nug}
Nugent P., et al, 1998, IAUC, 6804, 1

\bibitem[Pain et al(2002)]{pai}
Pain R., et al, 2002, ApJ accepted, astro-ph 0205476

\bibitem[Perlmutter et al(1997)]{per0}
Perlmutter S., et al, 1997, ApJ, 483, 565

\bibitem[Perlmutter et al(1999)]{per1}
Perlmutter S., et al, 1999, ApJ, 517, 565

\bibitem[Phillips(1993)]{phi0}
Phillips M. M., 1993, ApJ, 413, 105

\bibitem[Phillips et al(1999)]{phi}
Phillips M. M., Lira P., Suntzeff N. B., Schommer R. A.,
Hamuy M., Maza J., 1999, AJ, 118, 1766

\bibitem[Riess, Press \& Kirshner(1996)]{rpk}
Riess A. G., Press W. H., Kirshner R. P., 1996, ApJ, 473, 88

\bibitem[Riess et al(1998)]{rie1}
Riess A. G., et al, 1998, AJ, 116, 1009

\bibitem[Rowan-Robinson(2002)]{mrr}
Rowan-Robinson M., 2002, MNRAS, 332, 352

\bibitem[Saha et al(1999)]{sah}
Saha A., Sandage A., Tammann G. A., Labhardt L., Macchetto F. D.,
Panagia N., 1999, ApJ, 522, 802

\bibitem[Schmidt et al(1998)]{sch0}
Schmidt B., et al, 1998, ApJ, 507, 46

\bibitem[Schmidt(2000)]{sch}
Schmidt B., 2000, IAUC, 7516, 1

\bibitem[Tripp(1997)]{tri}
Tripp R., 1997, A\&A, 325, 871

\bibitem[Umeda et al(1999)]{ume1}
Umeda H., Nomoto K., Kobayashi C., Hachisu I., Kato M., 
1999, ApJ, 522L, 43

\bibitem[Wang et al(1997)]{wan}
Wang L., Hoeflich P., Wheeler J. C., 1997, ApJ, 483L, 29


\end{thebibliography}
\end{document}